\documentclass[prb,aps,showpacs,showkeys,groupedaddress,superscriptaddress,twocolumn,citeautoscript]{revtex4-2}

\usepackage[utf8]{inputenc}
\usepackage{color}
\usepackage{bbm}
\usepackage{amsfonts,amsmath,amssymb,stmaryrd}
\usepackage{graphicx}
\usepackage{subfigure}  
\usepackage{hyperref}
\usepackage{epsfig}
\usepackage{mathrsfs}
\usepackage{verbatim}
\usepackage{centernot}
\usepackage{hyperref}
\usepackage{color}
\usepackage{braket}

\newcommand{\tr}{\text{Tr}}

\renewcommand{\H}{\hat{\mathcal{H}}}
\renewcommand{\c}{\hat{c}}

\newcommand{\ct}{\tilde{c}}

\newcommand{\cd}{\hat{c}^\dagger}

\newcommand{\im}{\text{Im}}

\newcommand{\abs}[1]{\left\lvert{#1}\right\rvert}

\newcommand{\anticom}[2]{\lb #1, #2\rb}

\newcommand{\transpose}{\mathrm{T}}

\newcommand{\pd}[1]{\partial_{#1}}

\renewcommand{\d}{\mathrm{d}}

\newcommand{\gv}[2]{#1_{#2}}
\newcommand{\gvc}[2]{\overline{#1}_{#2}}
\newcommand{\gV}[1]{\vec{#1}}
\newcommand{\gVc}[1]{\vec{\overline{#1}}}
\renewcommand{\ol}[1]{\overline{#1}}
\newcommand{\lb}{\left\lbrace}
\newcommand{\rb}{\right\rbrace}
\newcommand{\la}{\left\langle}
\newcommand{\ra}{\right\rangle}
\newcommand{\lbr}{\left\lbrack}
\newcommand{\rbr}{\right\rbrack}

\usepackage{array}

\usepackage{bm}	
\renewcommand{\vec}[1]{\bm{#1}}

\begin{document}
\title{$\mathbb{Z}_2$ topological invariants for mixed states of fermions in time-reversal invariant band structures}

\author{Lukas Wawer}
\affiliation{Department of Physics and Research Center OPTIMAS, University of Kaiserslautern, 67663 Kaiserslautern, Germany}
\author{Michael Fleischhauer}
\affiliation{Department of Physics and Research Center OPTIMAS, University of Kaiserslautern, 67663 Kaiserslautern, Germany}


\begin{abstract}
The topological classification of fermion systems in mixed states is a long standing quest. For Gaussian  states, reminiscent of non-interacting unitary fermions, some progress has been made. While the topological quantization of certain observables such as the Hall conductivity is lost for mixed states, directly observable many-body correlators exist which preserve the quantized nature and naturally connect to known topological invariants in the ground state. For systems which break time-reversal (TR) symmetry, the ensemble geometric phase was identified as such an observable which can be used to define a Chern number in $(1+1)$ and $2$ dimensions. Here we propose a corresponding  $\mathbb{Z}_2$ topological invariant for systems with TR symmetry. We show that this mixed-state invariant is identical to well-known $\mathbb{Z}_2$ invariants for the ground state of the so-called fictitious Hamiltonian, which for thermal states is just the ground state of the system Hamiltonian itself. We illustrate our findings for finite-temperature states of a paradigmatic $\mathbb{Z}_2$ topological insulator, the Kane-Mele model.
\end{abstract}

\date{\today}
\maketitle


\section{Introduction}

Since the discovery of the quantum Hall effect \cite{Klitzing-PRL-1980,Tsui-PRL-1982,Laughlin-PRL-1983,Arovas-PRL-1984}, the topology of ground states of  many-body systems has become a key paradigm to classify phases of quantum matter \cite{Wen-RMP-2017}. Topological systems are characterized by integer-valued invariants \cite{TKNN-PRL-1982,Xiao-RMP-2010} which are responsible for characteristic features, such as quantized bulk transport or protected edge modes and their robustness to perturbations or deformations of the Hamiltonian. For systems of non-interacting fermions possible topological invariants can be fully classified by symmetries of the Hamiltonian under unitary and anti-unitary transformations in Fock space \cite{Altland-PRB-1997,Schnyder-PRB-2008,Ryu-NJPhys-2010,Kitaev}. 

An important question is whether the concept of topology, developed for quantum systems in the 
ground state, can be extended to finite temperatures or more general to non-equilibrium steady states of open systems, which both are described by density matrices \cite{Uhlmann-Rep-Math-Phys-1986,Bardyn-NJP-2013,Huang-PRL-2014,Viyuela-PRL-2014,Viyuela-PRL-2014b,Nieuwenburg-PRB-2014,Budich-Diehl-PRB-2015}. While a general solution to this problem is still open, some progress has been made for Gaussian states, which 
are fully characterized by the matrix of single-particle correlations. The symmetries of this matrix 
provide a topological classification according to the ten fundamental symmetry classes \cite{Altland2021}.
For finite-temperature states of 
fermions with Hamiltonians that break time-reversal (TR) symmetry, a consistent generalization that holds in 1+1 and 2 dimensions was given in  \cite{Bardyn-PRX-2018,Wawer-PRB-2021}. The same holds for non-equilibrium steady states that are Gaussian
\cite{Linzner-PRB-2016}. 

In the present paper we
propose a 
topological invariant for Gaussian mixed states of spinful fermions in $(1+1)$- and $2$-dimensional band structures with TR symmetry. This includes finite-temperature states of topological insulators such as the Kane-Mele (KM) model \cite{Kane-Mele-PRL-2005},
as well as non-equilibrium steady states with TR symmetric covariance matrix. We show that the mixed-state topological invariant is identical to the known $\mathbb{Z}_2$ invariant of the insulating ground state of the system Hamiltonian in the case of equilibrium states at any finite temperature or, respectively, of the ground state of the so-called fictitious Hamiltonian \cite{Bardyn-NJP-2013} directly related to the covariance matrix, in the case of non-equilibrium steady-states. 
We illustrate our findings with numerical simulations of the hallmark model system of a TR symmetric topological insulator, the KM model, at finite temperatures.

Insulators with TR breaking band structure, called Chern insulators  which include 
quantum Hall systems, have been the first for which the concept of topology was developed more than 35 years ago. In $2$ or $(1+1)$ spatial dimensions they are characterized by a topological invariant, the  Chern number. It can be written as an integral of the Berry curvature of occupied Bloch eigenstates $\ket{ u_n(\vec{k})}$ over the two- or $(1+1)$ dimensional Brillouin zone \cite{Xiao-RMP-2010}
\begin{eqnarray*}
C &=& i\sum_{n\,\mathrm{occup.}}\iint_{\mathrm{BZ}} \!\! \d k_x \, \d k_y \Bigl(\bigl\langle\partial_{k_x} u_n(\vec{k}) \bigr\vert \partial_{k_y} u_n(\vec{k})\bigr\rangle - \mathrm{c.c}\Bigr),
\end{eqnarray*}
or equivalently as the winding of the geometric Zak phase \cite{Zak1989}
\begin{eqnarray*}
 C = \frac{1}{2\pi} \int_{\mathrm{BZ}}\!\! \d k_y \frac{\partial \phi^\mathrm{Zak}_x(k_y)}{\partial k_y}=
-\frac{1}{2\pi} \int_{\mathrm{BZ}}\!\! \d k_x \frac{\partial \phi^\mathrm{Zak}_y(k_x)}{\partial k_x},
\end{eqnarray*}
where
\begin{eqnarray*}
\phi^\mathrm{Zak}_x(k_y) 
= i\int_{\mathrm{BZ}}\!\! \d k_x \, \langle u_n(\vec{k})\vert \partial_{k_x} u_n(\vec{k})\rangle,\quad \vec{k}=(k_x,k_y),
\end{eqnarray*}
in $x$ (or alternatively in $y$) direction. Using the relation between Zak phase and many-body polarization, 
$\Delta \phi^\mathrm{Zak}_\mu  = 2\pi \Delta P_\mu$, derived by King-Smith and Vanderbildt \cite{King-Smith-PRB-1993}, $C$ can alternatively be expressed as the winding of Resta's many-body polarization \cite{Resta-PRL-1998} in $x$ (or $y$) direction in the many-body ground state $\ket{ \Psi_0}$
\begin{eqnarray}
P_x(k_y)
& =& \frac{1}{2\pi}  \im\ln \bigl\langle \Psi_0\bigr\vert \hat T_x \bigl\vert \Psi_0\bigr\rangle.\label{eq:Resta}
\end{eqnarray}
Here $\hat T_x = e^{i\delta k_x \hat X(k_y)}$ is the momentum shift operator.  $\hat X(k_y)=\hat \Pi_{k_y} \hat X \hat \Pi_{k_y}$ is the projection of the position operator $\hat X$ on all single-particle states with momentum $k_y$ in $y$ direction and $\delta k_x=2\pi/L_x$ is the unit of lattice momentum 
in a system of $L_x\times L_y$ unit cells and periodic boundary conditions. 

The formulation of the Chern number in terms of expectation values via 
eq.\eqref{eq:Resta} allows for a straight-forward generalization to mixed states $\rho$. This leads to the \emph{ensemble geometric phase} (EGP):
\begin{equation}
    \phi^\mathrm{EGP}_{x,y} =  \im\ln \textrm{Tr}\Bigl\{ \rho\, \hat T_{x,y}\Bigr\}.\label{eq:EGP}
\end{equation}
In \cite{Bardyn-PRX-2018,Wawer-PRB-2021} we have shown that the winding of the EGP is a proper generalization of the Chern number to Gaussian mixed states, which is directly observable. It preserves the integrity of the topological invariant for equilibrium states and agrees with the ground state Chern number for any finite temperatures. It can change its value only if the energy gap of the Hamiltonian closes or for infinite temperatures. It also provides a valid  invariant to topologically classify Gaussian non-equilibrium states, provided that certain generalized gap conditions are fulfilled \cite{Bardyn-NJP-2013}. In this case it agrees with the Chern number of the ground state of the fictitious Hamiltonian, which has the same eigenstates as the covariance matrix.

In the early 2000th it was realized that also 2D (or 3D) band structures with TR symmetry can lead to topologically non-trivial insulating states for spinful fermions in systems with spin-orbit coupling \cite{Murakami-PRL-2004,Kane-Mele-PRL-2005,Fu-Kane-PRB-2006,Brenevig-PRL-2006,Koenig-Science-2007,Hazan-Kane-RMP-2010,Qi-RMP-2011}. These topological insulators have a bulk energy gap as any insulator but give rise to a quantized spin Hall effect and posses gapless edge modes protected by TR symmetry \cite{Murakami-PRL-2004,Kane-Mele-PRL-2005,Brenevig-PRL-2006,Koenig-Science-2007}. Due to TR symmetry the Chern number of quantum spin Hall systems vanishes and eigenstates come in pairs, called Kramers partners. As first shown by Kane and Mele \cite{Kane-Mele-PRL-2005} there is a $\mathbb{Z}_2$ invariant characterizing the topological properties of these systems. Several formulations for this  invariant have been proposed. In \cite{Kane-Mele-PRL-2005} Kane and Mele suggested the Pfaffian of the overlap matrix of Bloch wavefunctions and their Kramers partners, which was later shown to be equivalent to an expression based on the TR polarization by Fu and Kane \cite{Fu-Kane-PRB-2006}. An alternative, based on the winding of eigenvalues of the non-Abelian Wilson loop, was suggested by Yu et al. in \cite{Yu-PRB-2011}. A general classification of TR symmetric topological insulators in terms of dimensional reductions of a $(4+1)$ dimensional Chern-Simons field theory was given in \cite{Zhang2008}.

All of the above formulations of the $\mathbb{Z}_2$ invariant have in common that they require the knowledge of the single-particle Bloch functions of the Hamiltonian. This is in contrast to the TR-breaking case, for which the Chern number can be defined through an expectation value of a system-independent unitary operator $\hat T$, which is the basis of the generalization to density matrices via eq.\eqref{eq:EGP}. This makes the generalization of $\mathbb{Z}_2$ invariants to mixed states more complicated. A notable exception to the above mentioned $\mathbb{Z}_2$ indices is the difference of spin Chern numbers introduced in \cite{Sheng-PRL-2006}: For TR invariant systems where spin is conserved one can define separate Chern numbers for the individual spin  components $C_{\uparrow,\downarrow}$, which can be defined through expectation values of unitary operators. While their sum must vanish, their difference is a $\mathbb{Z}_2$ topological integer. It was shown in \cite{Sheng-PRL-2006} for the KM model that although the individual spin Chern numbers loose their meaning in the presence of spin-orbit coupling (SOC), their difference, called total spin Chern number, remains a suitable invariant. This number is however only a good topological index if the Hamiltonian with SOC can be adiabatically connected to a spin conserving Hamiltonian without closing the energy gap \cite{Fu-Kane-PRB-2006}.

Starting from the rather straightforward generalization of spin polarization and spin Chern numbers to finite-temperature states, applicable to Hamiltonians smoothly deformable to a spin-conserving limit, we introduce a generalization of the more general TR polarization to Gaussian mixed states of fermions for one- and two-dimensional band structures from which a mixed-state $\mathbb{Z}_2$  invariant can be constructed. 

The paper is organized as follows: In Sec.II we briefly discuss the hallmark model of a TR invariant topological insulator, the Kane-Mele model \cite{Kane-Mele-PRL-2005} and discuss the  formulations of $\mathbb{Z}_2$ topological invariants, relevant for our approach. In Sec.III we then present generalizations to mixed states both for the special case of thermal states of Hamiltonians that can be smoothly connected to a spin conserving limit and for the general case. A summary of our findings is given in Sec.IV.

\section{$\mathbb{Z}_2$ topological insulators}

\subsection{Kane-Mele model}

\begin{figure}[htb]
\includegraphics[width=0.18\textwidth]{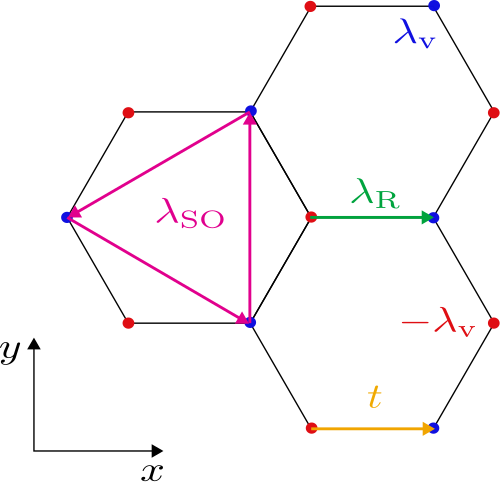} \qquad
\includegraphics[width=0.18\textwidth]{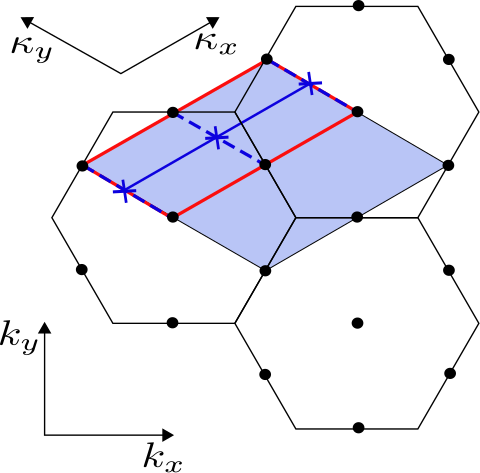} \qquad
\includegraphics[width=0.28\textwidth]{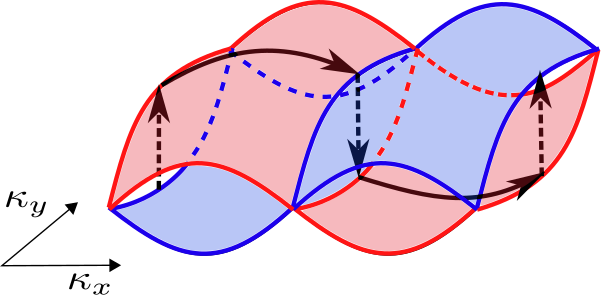}
\caption{(a) Kane-Mele model on Honeycomb lattice and all couplings between sites. (b) k-space representation with its TR points (black dots) and rotated unit cell (blue box). The blue line gives the path of the TR EGP and the blue crosses symbolize the band switching points. (c) Schematic path of TR EGP in the upper unit cell of Kane-Mele model (red box in b), adapted from Ref.\cite{Grusdt-PRA-2014}. Note that in $\kappa_y$ direction only half of the Brillouin zone is shown, i.e. $\kappa_y=[-\pi,0]$. The spectrum is repeated for $\kappa_y\in\{0,+\pi\}$, however with exchange of the color code that indicates Kramers bands I and II.}
\label{fig:Kane_Mele_model}
\end{figure}

A hallmark $\mathbb{Z}_2$ topological insulator is the Kane-Mele model \cite{Kane-Mele-PRL-2005}. It describes spin-full fermions on a 2D honeycomb lattice with nearest-neighbor (NN) and next-nearest neighbor (NNN) hopping and in the presence of Rashba spin-orbit coupling, see Fig.\ref{fig:Kane_Mele_model}. Its Hamiltonian
\begin{equation}
		\H_\mathrm{KM} = \H_\mathrm{H} + \H_\mathrm{R} +\H_\mathrm{v}
	\end{equation} 
contains three terms, where
	\begin{align}
	\H_\mathrm{H} &= t\!\! \sum_{\langle i,j\rangle,\sigma}\!\!\left( \cd_{i,\sigma} \, \c_{j,\sigma} + \mathrm{h.c.}\right)+i \lambda_\mathrm{SO}\!\!\! \sum_{\langle\langle i,j\rangle\rangle,\sigma}\!\! \nu_{ij} \,  \cd_{i,\sigma} \, \hat{s}^{z}_{\sigma,\sigma} \, \c_{j,\sigma}
	\end{align}
describes the graphene model. It contains spin-independent nearest neighbour $\langle i,j\rangle$ hopping and next-nearest neighbor $\langle\langle i,j\rangle\rangle$ hopping with spin-dependent sign, representing a spin-conserving SOC with strength $\lambda_{\mathrm{SO}}$. $\nu_{ij}=\pm1$ depending on whether the hopping makes a clockwise ($+$) or anti-clockwise ($-$) transition. The model has a unit cell of two sites and consists of two sublattices A and B. $\H_\mathrm{H}$ conserves TR symmetry and spin and opens a gap at the Dirac points of graphene such that a true insulator emerges. If a Rashba spin-orbit coupling 
	\begin{equation}
	\H_\mathrm{R} = i\lambda_\mathrm{R} \sum_{\langle i,j\rangle} \sum_{\sigma,\sigma^\prime} \cd_{i,\sigma}\, \left[\left(\vec{s}\times \vec{d}_{ij}\right)_z\right]_{\sigma,\sigma^\prime} \, \c_{j,\sigma^\prime}
	\end{equation}
is included, spin is no longer conserved. Here $\vec{d}_{ij}$ is the basis vector connecting site $i$ and $j$.  $\H_\mathrm{R}$ by itself still conserves TR symmetry and does not open the gap so that $\lambda_{\mathrm{SO}}\neq 0$ is necessary to get an insulator. The last term
	\begin{equation}
	\H_\mathrm{v} = \lambda_\mathrm{v} \sum_{j,\sigma} \epsilon_j \, \cd_{j,\sigma} \, \c_{j,\sigma}
	\end{equation}  
is added to	break inversion symmetry, where $\epsilon_j=\pm 1$ for site A or B respectively. 

As shown in \cite{Kane-Mele-PRL-2005} the Hamiltonian can be expressed in momentum-space as
	\begin{equation}
	\H_\mathrm{KM} = \sum_{\vec{k}} \vec{\ct}^\dagger(\vec{k})^\transpose \, \mathsf{h}_\mathrm{KM}(\vec{k}) \, \vec{\ct}(\vec{k}), \label{eq:Kane_Mele_k_space}
	\end{equation}
where we defined the vector of fermion annihilation and creation operators in momentum space as $\vec{\ct}(\vec{k})=\bigl(\ct_{\mathrm{A,\uparrow}},\ct_{\mathrm{B,\uparrow}},\ct_{\mathrm{A,\downarrow}},\ct_{\mathrm{B,\downarrow}}\bigr)^\transpose$ and the single particle Hamiltonian reads
\begin{equation}
	 \mathsf{h}_\mathrm{KM}(\vec{k})=\begin{pmatrix}
	 d_{\mathrm{SO},\uparrow} & d_t & 0 & \tilde{d}_{\mathrm{R}} \\
	 d_{t}^\star & -d_{\mathrm{SO},\uparrow}& d_{\mathrm{R}}&0 \\
	 0 & d_{\mathrm{R}}^\star & d_{\mathrm{SO},\downarrow} & d_t \\
	 \tilde{d}_{\mathrm{R}}^\star&0&d_t^\star & -d_{\mathrm{SO},\downarrow}
	 \end{pmatrix}
	\end{equation}
The coefficients entering this matrix 
	\begin{eqnarray*}
	d_{\mathrm{SO},\uparrow} &=& d_2 + d_{15}\\
	d_{\mathrm{SO},\downarrow} &=& d_2 - d_{15} \\
	d_t &=& d_1 + i d_{12} \\
	\tilde{d}_{\mathrm{R}} &=& -(d_4+d_{23})-i(d_3-d_{24}) \\
	d_{\mathrm{R}} &=& (d_4-d_{23})+i(d_3+d_{24})
	\end{eqnarray*}
are listed in table \ref{tab:Kane_Mele_coefficents}.
	
\begin{table}[htb]
		\centering
		\begin{tabular}{ccccc}
			\hline\hline
			$d_1$ & $t(1+2\cos \tilde k_x \cos \tilde k_y)$& & $d_{12}$ & $-2t \cos \tilde k_x \sin \tilde k_y$ \\
			$d_2$ & $\lambda_\mathrm{v}$& & $d_{15}$ & $\lambda_{\mathrm{SO}}(2 \sin 2\tilde k_x - 4 \sin \tilde k_x \cos \tilde k_y)$ \\
			$d_3$ & $\lambda_{\mathrm{R}}(1-\cos \tilde k_x \cos \tilde k_y)$ && $d_{23}$ & $-\lambda_{\mathrm{R}} \cos \tilde k_x \sin \tilde k_y$ \\
			$d_4$ & $-\sqrt{3}\lambda_{\mathrm{R}}\sin \tilde k_x \sin \tilde k_y$ && $d_{24}$ & $\sqrt{3} \lambda_{\mathrm{R}} \sin \tilde k_x \cos \tilde k_y$\\
			\hline\hline
		\end{tabular}
		\caption{Coefficients $d_\alpha$, $d_{\alpha,\beta}$ of the matrix elements of the KM model \eqref{eq:Kane_Mele_k_space} in k-space with $\tilde k_x = k_x a/2$ and $\tilde k_y = \sqrt{3} k_y a/2$, $a$ being the lattice constant. In the rotated coordinate system, see Fig.\ref{fig:Kane_Mele_model}b, one has
		$k_x=\frac{1}{2}(\kappa_x-\kappa_y)$ and $k_y=\frac{1}{2}(\kappa_x+\kappa_y)$}
		\label{tab:Kane_Mele_coefficents}
\end{table}
		
Without Rashba coupling $\lambda_{\mathrm{R}}=0$ there are in total four energy bands where each spin degree of freedom is gapped by $\Delta_\mathrm{gap} = \abs{6\sqrt{3}\lambda_{\mathrm{SO}}-2\lambda_\mathrm{v}}$ which separates the system in a spin quantum Hall (SQH) phase for $\lambda_{\mathrm{v}}<3\sqrt{3} \lambda_{\mathrm{SO}}$ and a trivial insulating phase  for $\lambda_{\mathrm{v}}>3\sqrt{3} \lambda_{\mathrm{SO}}$. The SQH phase is characterized by helical edge currents protected by TR symmetry, i.e. by opposite, clockwise respectively counterclockwise flowing currents of the two spin components. We note that although the Rashba term violates the spin conservation, there is a finite region $\lambda_{\mathrm{R}}<2\sqrt{3}\lambda_{\mathrm{SO}}$ for which the system can be smoothly connected to the SQH phase at $\lambda_{\mathrm{R}}=0$ without closing of a gap. In the rotated Brillouin zone, $\kappa_x \times \kappa_y = \lbr -\pi,\pi \rbr \times \lbr -\pi,\pi \rbr$, see Fig. \ref{fig:Kane_Mele_model}b, where $\vec{\kappa} =(\kappa_x,\kappa_y) = (k_x+k_y,k_y-k_x)$, there are nine TR points for $\kappa_x, \kappa_y \in \lb -\pi,0,\pi \rb$. Six of which are indicated in  Fig.\ref{fig:Kane_Mele_model}c. The KM model is TR symmetric,
\begin{equation}
    \sigma_y \mathsf{h}_\mathrm{KM}(\vec{k}) \sigma_y = 
    \mathsf{h}_\mathrm{KM}^*(-\vec{k})
\end{equation}
where $\sigma_y$ acts in spin space and, as for all TR-invariant Hamiltonians, has a vanishing Chern number $C=0$. 

An important property of Chern insulators with broken TR symmetry is that all single-particle Bloch states $\ket{u_n(\vec{k})}$ with energy below the Fermi level enter in exactly the same way in the expression of the topological invariant. No further knowledge about the Hamiltonian is needed to define the invariant. This is
no longer the case for systems with TR symmetry.
Here according to Kramers theorem every Bloch state at lattice momentum $\vec{k}$ is degenerate with a time-reversed Bloch state at $-\vec{k}$, called the Kramers partner.  Energy bands come in pairs and the definition of topological invariants requires to keep track of the Kramers partners separately. 

\subsection{Spin polarisation and spin Chern number}

For a two-dimensional lattice model of spinful fermions, such as the KM Hamiltonian \eqref{eq:Ham_real}, one can define separate Zak phases or polarizations in $x$ (or $y$) direction for the two spin components in the many-body ground state
\begin{equation}
    P_x^\sigma(k_y) = \frac{1}{2\pi}\im\ln \bigl\langle \Psi_0\bigr\vert \hat T_x^\sigma \bigl\vert \Psi_0\bigr\rangle,\qquad \sigma\in \{\uparrow,\downarrow\}.\label{eq:Psigma}
\end{equation}
The momentum shift operator $\hat T_x^\sigma = e^{i\delta k_x \hat X_\sigma(k_y)}$ now contains the position operator $\hat X_\sigma(k_y)$ projected  on both, a given value of $k_y$ and spin projection $\sigma$.

Without Rashba SOC, $\lambda_\mathrm{R}=0$, the KM Hamiltonian conserves the spin projection in $z$ direction. In an insulating state with equal spin populations, the windings of both spin polarizations, i.e the spin Chern numbers $C_\uparrow$ and $C_\downarrow$ are then individually quantized. Time-reversal symmetry dictates that their sum vanishes 
\begin{eqnarray}
C= C_\uparrow + C_\downarrow =\int_\mathrm{BZ}\!\!\! \d k_y \frac{\partial P_x^\uparrow(k_y)}{\partial k_y} + \frac{\partial P_x^\downarrow(k_y)}{\partial k_y} =0.
\end{eqnarray}
The difference modulo 2 is however a $\mathbb{Z}_2$ topological index
\begin{equation}
    C_\mathrm{sc} = \frac{1}{2}\bigl\vert C_\uparrow - C_\downarrow\bigr\vert.\label{eq:Spin-Chern} 
\end{equation}
If it is nonzero the system possesses conducting helical edge modes protected by TR symmetry. 

In the presence of Rashba SOC both spin components mix and the polarization windings of the individual spin components $C_\uparrow$ and $C_\downarrow$ are no longer quantized. To define a topological invariant for $(1+1)$ or 2-dimensional lattice systems without spin conservation Sheng et al. \cite{Sheng-PRL-2006} introduced the Chern number matrix by considering Hamiltonians with twisted boundary conditions and twist angles $\theta_x^\sigma$ and $\theta_y^\sigma$ in $x,y$ direction and for each spin component $\sigma$.
Then the $2\times 2$ Chern matrix
\begin{equation}
    C^{\alpha,\beta} = \frac{i}{4\pi} \iint \! \d \theta_x^\alpha \d \theta_y^\beta \left(\biggl\langle \frac{\partial \Psi_0}{\partial \theta_x^\alpha}\biggr\vert 
    \frac{\partial \Psi_0}{\partial \theta_y^\beta}\biggr\rangle -\mathrm{c.c.} \right)
\end{equation}
is a gauge-invariant quantity, which has the advantage that its definition does not require any knowledge about the eigenstates of the Hamiltonian. The authors then argued that the spin related Chern number (here with an additional factor of $1/2$)
\begin{equation}
    C_\mathrm{sc} = \frac{1}{2}\sum_{\alpha,\beta} \alpha C^{\alpha,\beta} \label{eq:C_sc}
\end{equation}
generalizes the $\mathbb{Z}_2$ topological invariant of Fu, Kane and Mele to one with three values $0,\pm 1$. It was shown later, however, that the two cases $\pm 1$ characterize the same topological phase. The spin Chern number \eqref{eq:C_sc} is however only a good topological index if the Hamiltonian can be adiabatically connected to one with conserved spin components \cite{Fu-Kane-PRB-2006}.

\subsection{TR polarization}

A more general definition of a topological invariant for TR symmetric band structures is based on the so-called TR polarization. 
Due to TR symmetry every band splits into two subbands of Kramers pairs labeled I,II whose Bloch wavefunctions are related by time reversal 
\begin{equation}
    \ket{ u_\mathrm{II} (-\vec{k})}  = e^{i\chi(\vec{k})}  {\cal T} \ket{ u_\mathrm{I} (\vec{k})}.
\end{equation}
Here ${\cal T}= i\sigma_y {\cal K}$ is the anti-unitary TR operator, where ${\cal K}$ is complex conjugation and $\sigma_y$ acts in spin space. Fu and Kane introduced the TR polarization (e.g. in $x$ direction) as the difference of the polarizations or Zak phases of Kramers pairs for TR lattice momenta in $y$ direction $\kappa_y=0,\pi$, for which the subbands cross at $\kappa_x=0,\pm \pi$
\begin{equation}
    P_\theta(\kappa_y) = P^\mathrm{I}(\kappa_y) - P^\mathrm{II}(\kappa_y),\quad \mathrm{for}\quad \kappa_y=0,\pi\label{eq:TR-old}
\end{equation}
where
\begin{equation}
    P^\mathrm{I}(\kappa_y)= \frac{i}{2\pi} \int_\mathrm{BZ} \!\! \d\kappa_x \, \langle u^\mathrm{I}(\vec{\kappa}) \vert \partial_{\kappa_x} u^\mathrm{I}(\vec{\kappa})\rangle\label{eq:PI}
\end{equation}
and similarly for $P^\textrm{II}$. For the following discussion it is useful to consider the discretized version of eq.\eqref{eq:PI}, also used in numerical implementations. Using $\kappa_x = m \delta\kappa_x$, with $\delta\kappa_x=2\pi/L_x$, $N=L_x\times L_y$ being the number of unit cells in a system with periodic boundary conditions
\begin{eqnarray}
 P^\textrm{I} &=& \mathrm{arg}\exp\left\{ -\frac{i}{2\pi} 
 \int\! \d\kappa_x \, \Bigl\langle \partial_{\kappa_x} u^\mathrm{I}(\vec{\kappa})  \Bigr\vert  u^\mathrm{I}(\vec{\kappa})\Bigr\rangle
 \right\}\biggr\vert_{\kappa_y=0,\pi}\nonumber\\
 &=& \frac{1}{2\pi}\mathrm{arg} \prod_{m=0}^{L-1} \bigl\langle u^\textrm{I}(\kappa_x + \delta\kappa_x,\kappa_y)\bigr\vert u^\mathrm{I}(\kappa_x,\kappa_y)\bigr\rangle\biggr\vert_{\kappa_y=0,\pi},
\end{eqnarray}
where $\ket{ u^\textrm{I}(2\pi,\kappa_y)} = \ket{ u^\textrm{I}(0,\kappa_y)}$. The difference of $P_\theta$ at the two TR momenta $\kappa_y=0,\pi$ then defines a  $\mathbb{Z}_2$ topological invariant, provided the same continuous gauge has been used for the TR polarizations
\begin{equation}
    \nu_2 = P_\theta(\pi) - P_\theta(0)\quad \mathrm{mod}\, 2\, .\label{eq:nu2}
\end{equation}
The same continuous gauge is necessary since otherwise $P_\theta(\pi)$ and $P_\theta(0)$ could independently be changed by gauge transformations and $\nu_2$ would no longer unambiguously be defined. 

\subsection{Continuous TR (cTR) polarization}
\label{cTR}

The requirement of a continuous gauge, needed for the definition of $\nu_2$ in \eqref{eq:nu2}, poses a challenge for the generalization to mixed states. For this it would be much more convenient to define the $\mathbb{Z}_2$ invariant as a winding number along a
continuous path $\kappa_y =0 \to \pi$. The problem here is that $P_\theta(\kappa_y)$ is discontinuous at the TR lattice momenta $\kappa_y=0,\pm \pi$. This is because for $\kappa_y=0,\pi$ the Kramers partners $\ket{ u^\textrm{I,II}(\vec{\kappa})}$ exchange their energetic order when crossing the degeneracy point at $\kappa_x=0$, i.e when going from the first half of the Billouin zone, $\kappa_x=[-\pi,0)$, to the second, $\kappa_x=(0,\pi]$. As indicated in Fig.\ref{fig:Kane_Mele_model}c the degeneracy is in general lifted when going away from $\kappa_y=0,\pm\pi$. 

A continuous version of the TR polarization has been introduced in \cite{Grusdt-PRA-2014}. Here in the definition of the polarization or Zak phase, subbands are switched in the integral over $\kappa_x$ at $\kappa_x=0$. This is indicated by the black arrows in Fig.\ref{fig:Kane_Mele_model}c. This leads to the modified partner polarizations
\begin{eqnarray}
P^\mathrm{i}(\kappa_y) =  \frac{1}{2\pi}\mathrm{arg}&&\!\!
 \prod_{\kappa_x=-\pi+\delta \kappa_x}^{-\delta \kappa_x} \! \bigl\langle u^{u}(\kappa_x + \delta\kappa_x)\bigr\vert u^{u}(\kappa_x)\bigr\rangle \nonumber\\
&& \quad \times\langle u^l(\delta \kappa_x)\vert u^u(0)\rangle\\
 \quad \times&&\! \prod_{\kappa_x=\delta \kappa_x}^{\pi-\delta \kappa_x} \!
\bigl\langle u^{l}(\kappa_x + \delta\kappa_x)\bigr\vert u^{l}(\kappa_x)\bigr\rangle
\nonumber\\
&&\quad \times \langle u^u(-\pi+\delta \kappa_x)\vert u^l(\pi)\rangle\nonumber
\end{eqnarray}
where we suppressed the dependence on $\kappa_y$, and $\vert u^{u,l}(\vec{\kappa})\rangle$ refers to the energetically upper ($u$) and lower ($l$) subband, see Fig.\ref{fig:Kane_Mele_model}c. The second partial Polarization $P^\textrm{ii}$ is defined analogously. The corresponding continuous TR (cTR) polarization is denoted as 
\begin{equation}
    \widetilde P_\theta(\kappa_y) = P^\mathrm{i}(\kappa_y) - P^\mathrm{ii}(\kappa_y).
\end{equation}
Since $P^\mathrm{i,ii}(\kappa_y)$ are now smooth in $\kappa_y$ the $\mathbb{Z}_2$ invariant can be defined as a winding number over half the Brillouin zone in $y$ direction
\begin{equation}
    \nu_2 = \int_{0}^\pi\!\! \d \kappa_y \frac{\partial}{\partial \kappa_y} \widetilde P_\theta(\kappa_y).
\end{equation}

\section{$\mathbb{Z}_2$ invariants for Gaussian mixed states}

\subsection{Gaussian states}

We here want to consider Gaussian density matrices of fermions, which are the mixed-state analog of ground states of non-interacting particles. Furthermore we restrict ourselves to states that commute with the total number operator. The latter is not necessary but it makes the discussion more transparent and includes the most interesting cases. Gaussian density matrices have the form 
\begin{equation}
\hat{\rho} \sim \, \exp\Bigl(-\sum_{i,j} \cd_i \, \mathsf{G}_{ij} \, \c_j\Bigr)\label{eq:Gauss}
\end{equation}
and are fully determined by the matrix $G_{ij}$, which defines a \emph{fictitious Hamiltonian} \cite{Bardyn-NJP-2013}
\begin{equation}
\H_\mathrm{fict} = \sum_{i,j} \cd_i \, \mathsf{G}_{ij} \, \c_j \label{eq:Ham_fict}
\end{equation}
and whose elements are given by single-particle correlations
\begin{equation}
\langle\cd_i\,\c_j\rangle =\left[f(\mathsf{G})\right]_{ji}=\frac{1}{2}\lbr\mathbbm{1} - \tanh \left(\frac{\mathsf{G}}{2}\right)\rbr_{ji}. \label{eq:correlations}
\end{equation}
Gaussian mixed states result e.g. as steady states of systems coupled to Markovian reservoirs that are described by Lindblad master equations with quadratic Hamiltonians and Linblad generators linear in particle creation and annihilation operators. Also thermal states of non-interacting (and particle number conserving) fermion Hamiltonians in a canonical ensemble are Gaussian. Here
\begin{equation}
\hat{\rho} = \frac{1}{\mathcal{Z}} \, e^{-\beta \, \H} \label{eq:thermal_state}
\end{equation}
with partition function $\mathcal{Z} = \la\exp\left(\beta \H\right)\ra$ and the real-space Hamiltonian 
\begin{equation}
\H = \sum_{i,j} \cd_i \, \mathsf{h}_{ij} \, \c_j. \label{eq:Ham_real}
\end{equation}
Gaussian states have the advantage that they allow for largely analytic evaluations of expectation values. In Grassmann representation the expectation value of normal-ordered operators $\hat{O}$ can be written as
\begin{equation}
\bigl\langle\hat{O}\left(\mathbf{\cd},\mathbf{\c}\right)\bigr\rangle = \det\left(f(\mathsf{G})\right) \int \d\left(\vec{\psi},\vec{\ol{\psi}}\right) \, e^{\vec{\ol{\psi}} \, f(\mathsf{G})^{-1} \, \vec{\psi}} \, O\left(\vec{\ol{\psi}},\vec{\psi}\right). \label{eq:Gaussian_Grassmann_operator}
\end{equation}
Here $O\left(\vec{\ol{\psi}},\vec{\psi}\right)$ is obtained by replacing operators in $\hat{O}\left(\mathbf{\cd},\mathbf{\c}\right)$ by Grassmann numbers, $\cd_i \to \ol{\psi}_i$ and $\c_i\to \psi_i$. The function $f(\mathsf{G})=\left(e^\mathsf{G}+\mathbbm{1}\right)^{-1}$ is directly linked to the fictitious Hamiltonian, eq.\eqref{eq:Ham_fict}.

\subsection{Spin polarization for mixed states}

\begin{figure}[htb]
\centering
\includegraphics[width=0.48\textwidth]{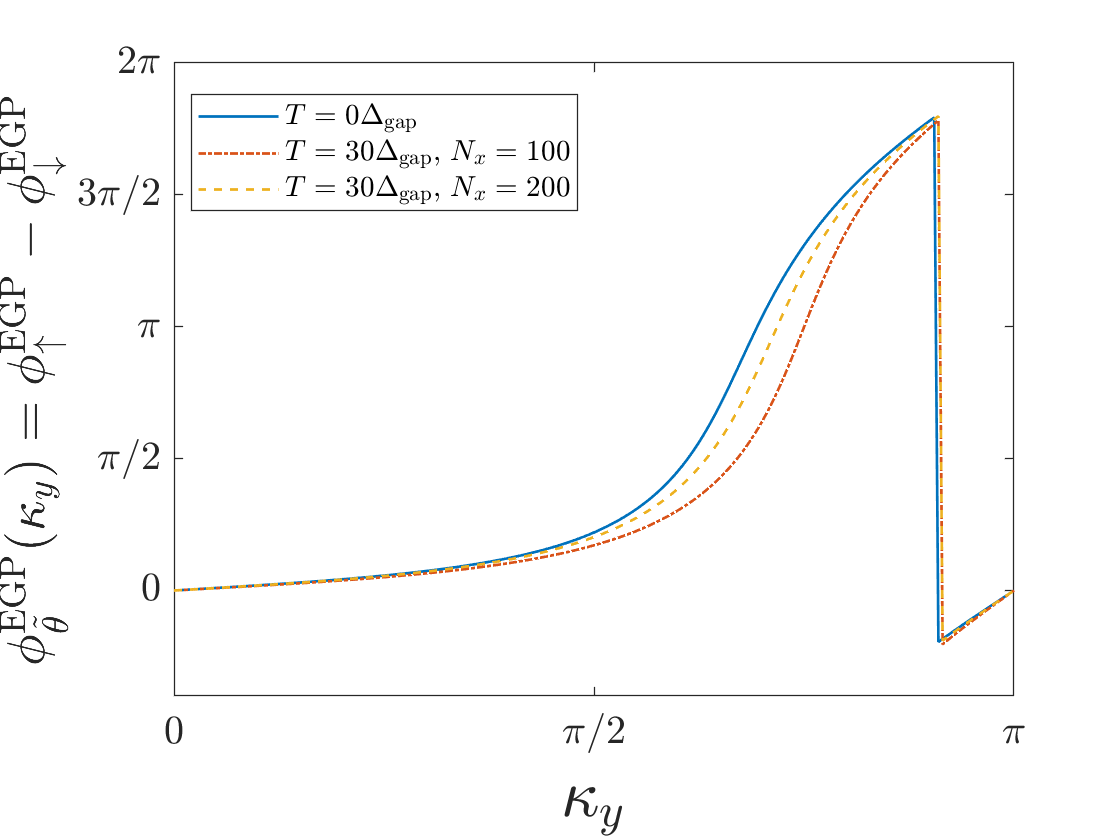}
\includegraphics[width=0.48\textwidth]{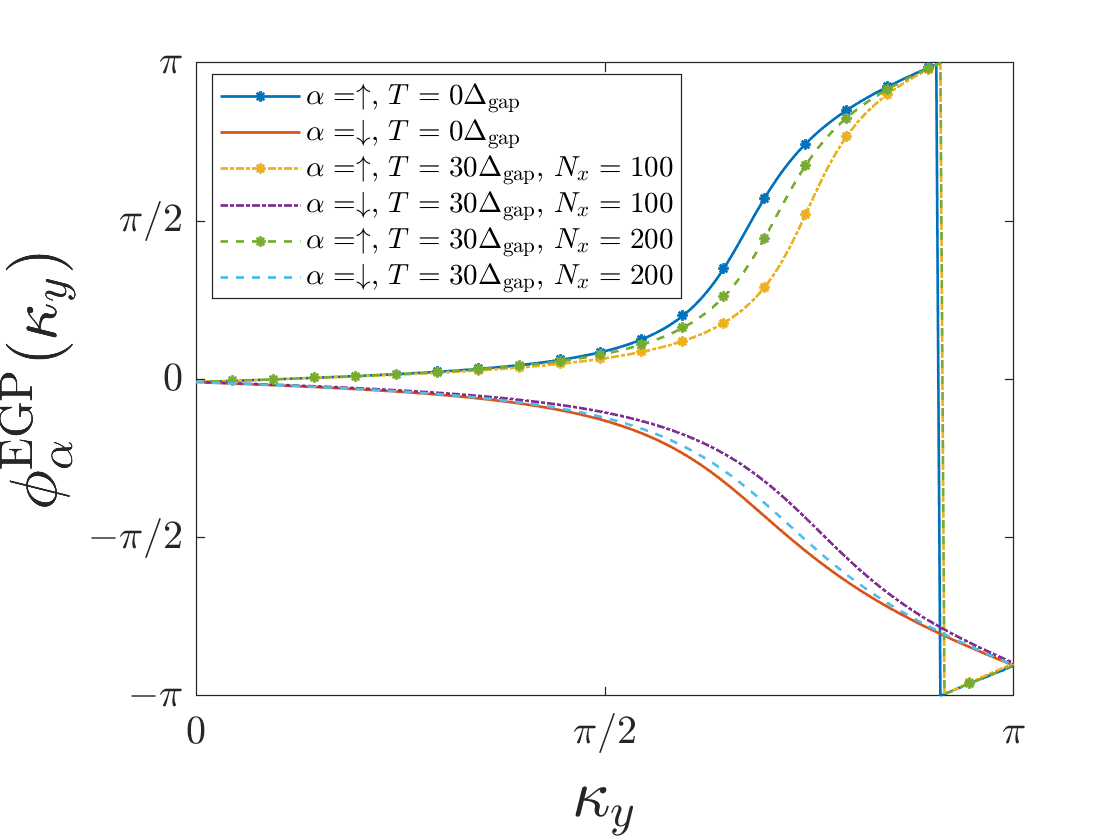}
\caption{TR spin EGP $\phi_{\tilde{\theta}}^\mathrm{EGP}(\kappa_y)=\phi_{\uparrow}^\mathrm{EGP}(\kappa_y)-\phi_\downarrow^\mathrm{EGP}(\kappa_y)$ (top) and Kramer's pair EGP $\phi_\alpha^\mathrm{EGP}(\kappa_y)$ for $\alpha = \uparrow,\downarrow$ (bottom) of Kane-Mele model.
Parameter of the model are $\lambda_\textrm{SO}=0.06 t$, $\lambda_\textrm{R}=0.05 t$, and $\lambda_\textrm{v}=0.1 t$.}
\label{fig:TR_spin_EGP}
\end{figure}

The spin polarization, defined in eq.\eqref{eq:Psigma}, can straightforwardly be generalized to density matrices in full analogy to Refs.\cite{Linzner-PRB-2016,Bardyn-PRX-2018} leading to spin version of the EGP
\begin{equation}
    \phi_{\sigma}^\mathrm{EGP}(\kappa_y) = \im\ln \mathrm{Tr}\Bigl\{\rho\, \hat T_x^\sigma\Bigr\}.
\end{equation}
Following exactly the same reasoning as in \cite{Bardyn-PRX-2018}, we realize that in a finite-temperature state $\rho$ the TR spin EGP
\begin{eqnarray}
 \phi_{\theta}^\mathrm{EGP}(\kappa_y) =  \phi_{\uparrow}^\mathrm{EGP}(\kappa_y)-\phi_{\downarrow}^\mathrm{EGP}(\kappa_y)
\end{eqnarray}
approaches its value at $T=0$ in the thermodynamic limit of infinite system size. In Fig.\ref{fig:TR_spin_EGP} we have plotted $\phi_{\theta}^\textrm{EGP}$ as well as the individual spin EGPs $\phi_{\uparrow}^\textrm{EGP}$ and $\phi_{\downarrow}^\textrm{EGP}$ as function of $\kappa_y$ for the KM model. Shown are the curves for the ground state ($T=0$) and for a thermal state at a high temperature, much above the single-particle energy gap ($T=30\Delta_\mathrm{gap}$) for two different system sizes. One clearly recognizes that all EGPs approach the corresponding $T=0$ values. Furthermore irrespective of the system size the windings over half the Brillouin zone, which define the $\mathbb{Z}_2$ topological invariant, are always the same, i.e. the topological index is independent of system size identical to the ground state value. 

\subsection{cTR polarization for mixed states}

While the generalization of a $\mathbb{Z}_2$ topological invariant to thermal states of Hamiltonians that can adiabatically be deformed to spin conserving Hamiltonians is straight forward, the general case is much more involved. This is because the TR polarization, eq.\eqref{eq:TR-old}, cannot directly be expressed in terms of an expectation value of a unitary operator. Instead one has to keep track of Kramers partners, which requires knowledge of the eigenstates of the fictitious Hamiltonian $\sf G(\vec{k})$. In order to construct a mixed-state $\mathbb{Z}_2$ topological invariant we thus employ the cTR polarization introduced in
Sec.\ref{cTR} and make use of its Grassmann representation.

\subsubsection{Grassmann representation of EGP and gauge reduction}

Let us begin by recapitulating the Grassmann representation of the \emph{single}-species ensemble geometric phase, eq.\eqref{eq:EGP}. For the EGP we have to determine the expectation value of the momentum shift operator $\braket{\hat{T}}$ in a Gaussian state. 
A normal-ordered form of the momentum shift operator $\hat{T}(\vec{\cd},\vec{\c})$ in real space can be obtained noting
\begin{align}
\hat{T} = \hat{T}(\vec{\cd},\vec{\c})&=e^{i\delta k\sum_j x_j \hat{n}_j} \nonumber\\
&= \prod_{j}\left(1+\left(e^{i\delta k x_j}-1\right)\hat{n}_j\right). \label{eq:momentum_shift_fermion}
\end{align}
Replacing $\cd_j \to \ol{\psi}_j$ and $\c_j\to \psi_j$, where $j=(r,s)\in[1,L]\times[1,M]$ is a real-space multi index for $L$ unit cells and $M$ subbands corresponding to $M$ lattice sites per unit cell, results in the Grassmann representation
\begin{align}
T\left(\vec{\overline{\psi}},\vec{\psi}\right) &= \prod_{j}\left(1+\left(t_j-1\right)\gvc{\psi}{j}\gv{\psi}{j}\right) \nonumber\\
&= \exp\lb \sum_j \gvc{\psi}{j}\left(t_j-1\right) \gv{\psi}{j} \rb \nonumber\\
&= \exp\Bigl\{\gVc{\psi}\left(\mathsf{T}-\mathbbm{1}\right)\gV{\psi}\Bigr\} \label{eq:T-Grassmann}
\end{align}
$\mathsf{T}=\mathrm{diag}(t_j)$ is the diagonal matrix of the single momentum shifts $t_j=e^{i\delta k x_j}$. 

In order to evaluate the expectation value of $\hat {T}$ in a translation invariant Gaussian state \eqref{eq:Gauss}, we first go to momentum space $\psi_j\to\tilde\psi(k)$, where the fictitious Hamiltonian ${\sf G}$ factorizes into $M\times M$ matrices $ \tilde{\sf G}(k)$.
In $k$-space the momentum shift matrix becomes block diagonal in the first lower minor diagonal
\begin{equation}
\tilde{\mathsf{T}} =\begin{pmatrix}
	0& & &\mathbbm{1}_M \\
	\mathbbm{1}_M&0& & \\
	&\ddots & \ddots & \\
	&&\mathbbm{1}_M&0
	\end{pmatrix}.
\end{equation}
We then transform into an eigenbasis of $\tilde{\sf G}(k)$ by a unitary transformation 
$\tilde{\mathsf{B}}(k)=\mathsf{U}^\dagger \tilde{\sf G}(k)\mathsf{U}$ such that  
$\tilde{\mathsf{B}}(k)=\mathrm{diag}_{(k,s)}(\tilde{\beta}_{k,s})=\mathrm{diag}_{(k,s)}({\beta}\varepsilon_{k,s})$ being the eigenvalue spectrum of the fictitious Hamiltonian (called purity spectrum). Here we used the notation ${\beta}\varepsilon_{k,s}$ as the fictitious Hamiltonian of a thermal state is just the true Hamiltonian multiplied by the inverse temperature $\beta=1/k_B T$. The unitary matrix $\mathsf{U}$ is block-diagonal in $k$-space
\begin{equation}
    \mathsf{U}= \mathrm{diag}_k \mathsf{U}_k
\end{equation}
with $\mathsf{U}_k$ being an $M\times M$ matrix.
This then yields with $\tilde \psi(k) = U \phi(k)$
\begin{align}
\langle \hat T\rangle &= \det\left(f(\tilde{\mathsf{B}})\right) \int \d\left(\vec{{\overline{\phi}}},\vec{{\phi}}\right) \, e^{\vec{{\overline{\phi}}}\left(f({\mathsf{B}})^{-1}-\mathbbm{1}+{\mathsf{U}}^\dagger \tilde{\mathsf{T}}{\mathsf{U}}\right)\vec{{\phi}}} \nonumber\\
&= \det\left(\mathbbm{1}-f(\tilde{\mathsf{B}})+f(\tilde{\mathsf{B}}) \, {\mathsf{U}}^\dagger \tilde{\mathsf{T}}{\mathsf{U}}\right)\\
&=\det\left(\mathbbm{1}-f(\tilde{\mathsf{B}})\right) \det\left(\mathbbm{1}+e^{-\tilde{\mathsf{B}}} \, {\mathsf{U}}^\dagger \tilde{\mathsf{T}}{\mathsf{U}}\right).\nonumber
\label{eq:T_Grassmann}
\end{align}
The argument of this expression then gives the EGP $\phi^\mathrm{EGP}= \mathrm{arg}\left(\langle \hat T\rangle\right)$.It is straight forward to show that at zero temperature $f(\tilde{\mathsf{B}})=\bigl(e^{\tilde{\mathsf{B}}}+1\bigr)^{-1}$ is unity for all occupied bands and vanishes otherwise. If there is only one occupied band this gives
\begin{align}
\phi^\mathrm{EGP} &= \arg \det\left(\lbr{\mathsf{U}}^\dagger \tilde{\mathsf{T}}{\mathsf{U}}\rbr_{(k,0),(k^\prime,0)}\right) \nonumber \\
&=\arg \prod_{k} \langle{u_{0}(k+\delta k)}\vert u_{0}(k)\rangle = \phi^\mathrm{Zak}_0.
\end{align}
In \cite{Bardyn-PRX-2018} we have shown that the EGP approaches the ground state Zak phase in the thermodynamic limit $L\to\infty$. This can be seen from \eqref{eq:T_Grassmann} using $\det(\cdots) = \exp\mathrm{Tr}\log(\cdots)$ and the series expansion of $\log(\mathbbm{1}+\mathsf{A})$ where $\mathsf{A}=e^{-\tilde{\mathsf{B}}}\, {\mathsf{U}}^\dagger\tilde{\mathsf{T}}{\mathsf{U}}$ and $\mathsf{A}^L=(\prod_k \mathsf{A}_{k})\otimes \mathbbm{1}$ to obtain 
\begin{align}
\phi^\mathrm{EGP} &= \arg \det\left(\mathbbm{1}+(-1)^{L+1} \prod_k \mathsf{A}_{k} \right) \nonumber\\
&= \arg \det\left(\mathbbm{1}+(-1)^{L+1} \prod_k e^{-\tilde{\mathsf{B}}_k}\, {\mathsf{U}}_{k+\delta k}^\dagger {\mathsf{U}}_k\right) \nonumber \\
&= \arg \det\left(\mathbbm{1}+\tilde{\mathsf{M}}_T\right) \label{eq:EGP_gauge_reduction}
\end{align}
where in the last line we introduced the abbreviation $\tilde{\mathsf{M}}_T$ for the path-ordered matrix product. In the thermodynamic limit the latter is a product of $M\times M$ link matrices ${\mathsf{U}}_{k+\delta k}^\dagger {\mathsf{U}}_k$ and weighting factors $e^{-\tilde{\mathsf{B}}_k}$. The matrix elements of the link matrices
\begin{eqnarray}
 &&\langle q,s\vert {\mathsf{U}}_{k+\delta k}^\dagger {\mathsf{U}}_k \vert q^\prime,s^\prime\rangle = \langle k+\delta k,s\vert k, s^\prime\rangle \delta_{q,k+\delta k} \delta_{q^\prime,k} \\
 &&\qquad \approx 1 - \delta k \langle k,s\vert \partial_k\vert k, s^\prime\rangle\approx  \exp\left\{i\delta k \mathcal{A}_{s,s^\prime}(k)\right\} \nonumber
\end{eqnarray}
are just phase factors containing the non-Abelian Berry connection $\mathcal{A}_{s,s^\prime}(k)$. For mixed states with a purity gap the real weighting factors $e^{-\tilde{\mathsf{B}}_k}$ will select the purity bands with the lowest eigenvalue corresponding to the many-body ground state of the fictitious Hamiltonian in the thermodynamic limit, where the number of terms in the product becomes infinite \cite{Bardyn-PRX-2018}. As a consequence the EGP approaches the Zak phase of the ground state of the fictitious Hamiltonian.

\subsubsection{Grassmann representation of cTR polarization}

We now want to extend the above discussion to fermions with \emph{two} spin components and find an expression for the cTR polarization. In this case each orbital band splits into two. The total momentum shift operator that shifts both spin components then reads in generalization of eq.\eqref{eq:T-Grassmann} in Grassmann representation
\begin{align}
	\tilde{T}\left(\vec{{\overline{\phi}}},\vec{{\phi}}\right) 	&=\exp \lb\vec{{\overline{\phi}}}\left({\mathsf{U}}^\dagger\tilde{\mathsf{T}}{\mathsf{U}}-\mathbbm{1}\right)\vec{{\phi}} \rb \nonumber\\
	&= \exp \lb\vec{{\overline{\phi}}}\left({\mathsf{X}}-\mathbbm{1}\right)\vec{{\phi}}\rb
\end{align}
where we used the basis $\vec{{\phi}}=(\vec{{\phi}}^u,\vec{{\phi}}^l)$ and the link matrices ${\mathsf{X}}_{k_x,k_x^\prime} = {\mathsf{X}}_{k_x}(k_y) \, \delta_{k_x+\delta k_x,k_x^\prime}$, where
	\begin{align}
	  {\mathsf{X}}_{k_x}(k_y)&= \begin{pmatrix}
	{\mathsf{U}}_{k_x+\delta k_x}^u (k_y){\mathsf{U}}_{k_x}^u (k_y) & {\mathsf{U}}_{k_x+\delta k_x}^u (k_y){\mathsf{U}}_{k_x}^l (k_y)\\
	{\mathsf{U}}_{k_x+\delta k_x}^l (k_y){\mathsf{U}}_{k_x}^u (k_y) & {\mathsf{U}}_{k_x+\delta k_x}^l(k_y){\mathsf{U}}_{k_x}^l(k_y)
	\end{pmatrix} \nonumber \\
	&= \begin{pmatrix}
	e^{i\mathcal{A}^{u,u}(\vec{k})} & e^{i\mathcal{A}^{u,l}(\vec{k})} \\
	e^{i\mathcal{A}^{l,u}(\vec{k})} & e^{i\mathcal{A}^{l,l}(\vec{k})}
	\end{pmatrix}
	\end{align}
where we used the non-Abelian Berry connection
	\begin{equation}
	\mathcal{A}^{s,s^\prime}_{\alpha,\alpha^\prime}(\vec{k}) = i \langle u^s_\alpha(\vec{k})\vert \pd{k_x} u^{s^\prime}_{\alpha^\prime}(\vec{k})\rangle.
	\end{equation}

From the above expressions we can see which modifications are needed to obtain the cTR polarization of Sec.\ref{cTR}. The link matrices ${\mathsf{X}}_{k_x}(k_y)$ should be projected to one subband $(i)$ or $(ii)$, which should be (say) the upper band for $k_x$ in one half of the Brillouin zone and the lower band in the second half of the Brillouin zone, so
\begin{equation}
    {\mathsf{X}}_{k_x}(k_y) \, \longrightarrow\, \tilde{\mathsf{X}}_{k_x}^i(k_y)
\end{equation}
where
\begin{equation}
    \tilde{\mathsf{X}}_{k_x}^{i}(k_y) = \begin{cases} {\mathsf{U}}_{k_x+\delta k_x}^u(k_y){\mathsf{U}}_{k_x}^u(k_y) \quad\mathrm{for}\quad k_x \in (-\pi,0) \\
    {\mathsf{U}}_{k_x+\delta k_x}^l(k_y){\mathsf{U}}_{k_x}^u(k_y) \quad\mathrm{for}\quad k_x =0 \\
    {\mathsf{U}}_{k_x+\delta k_x}^l(k_y){\mathsf{U}}_{k_x}^l(k_y) \quad\mathrm{for}\quad k_x \in (0,\pi) \\
    {\mathsf{U}}_{k_x+\delta k_x}^u(k_y){\mathsf{U}}_{k_x}^l(k_y) \quad\mathrm{for}\quad k_x =\pi
    \end{cases}
\end{equation}

%
%
%
Similarly we can define the link matrices for the momentum shift operators of band $(ii)$. With this we arrive at the Grassmann representation of the momentum shift operators for $s=i,ii$ 
\begin{align}
	\tilde{T}_s\left(\vec{{\overline{\phi}}},\vec{{\phi}}\right) &=
	 \exp \Bigl\{\vec{{\overline{\phi}}}\left(\tilde{\mathsf{X}}^s-\mathbbm{1}\right)\vec{{\phi}}\Bigr\}
\end{align}
such that the partial EGPs for both partners read
\begin{equation}
\phi^\mathrm{EGP}_s(k_y) = \arg \big\langle\tilde{T}_s\left(\vec{{\overline{\phi}}}_s,\vec{{\phi}}_s\right)\big\rangle.
\end{equation}
With this we can finally define the time-reversal EGP
	\begin{equation}
	\phi^\mathrm{EGP}_\theta(k_y) = \phi^\mathrm{EGP}_i(k_y) -\phi^\mathrm{EGP}_{ii}(k_y).
	\end{equation} 
The winding of $\phi^\mathrm{EGP}_\theta(k_y)$ over half the Brillouin zone then defines a general $\mathbb{Z}_2$ topological invariant
\begin{eqnarray}
 \nu_2^\textrm{EGP} = \frac{1}{2\pi}\int_{0}^\pi\!\! \d k_y \frac{\partial}{\partial k_y} \phi^\mathrm{EGP}_\theta(k_y).
\end{eqnarray}
Due to the gauge reduction mechanism outlined above for the TR broken case, one can show that the time-reversal EGP approaches the value in the ground state of the Hamiltonian, respectively the fictitious Hamiltonian, in the thermodynamic limit. The topological invariant 
is the same for all system sizes.

\subsubsection{$\mathbb{Z}_2$ topological invariant for Kane-Mele model at finite temperature}

\begin{figure}[htb]
\centering
\includegraphics[width=0.48\textwidth]{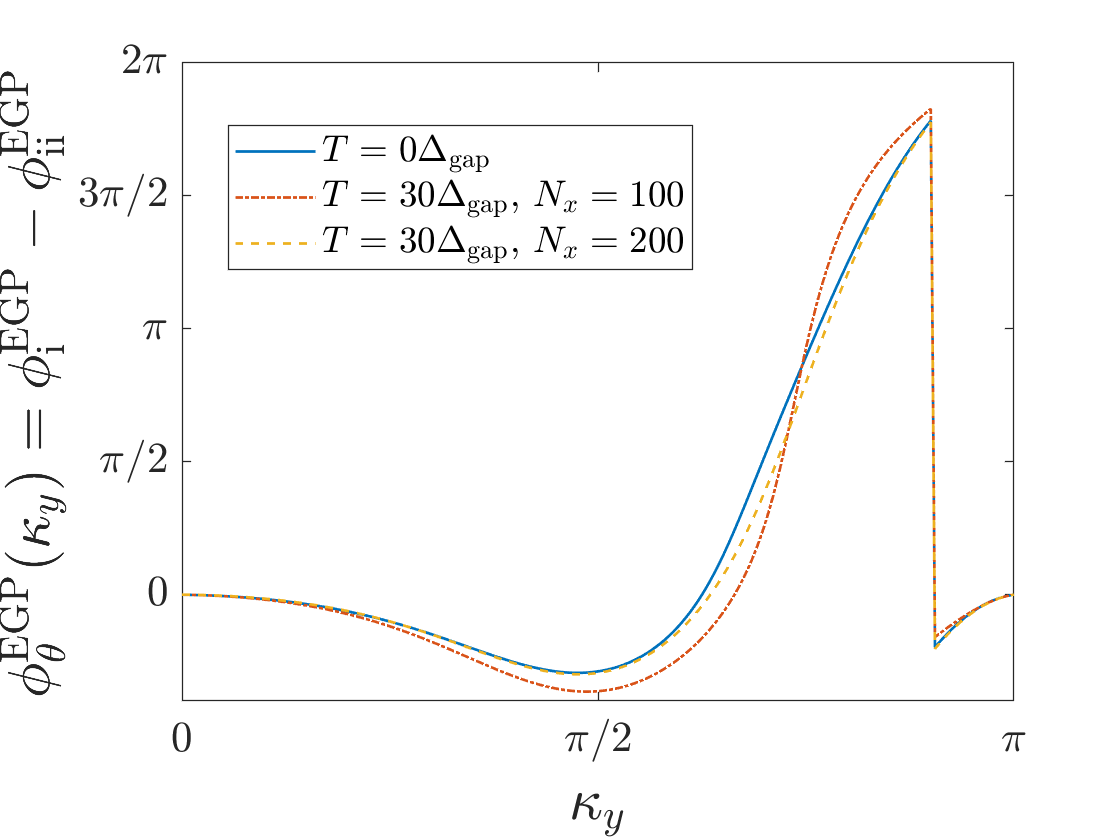}
\includegraphics[width=0.48\textwidth]{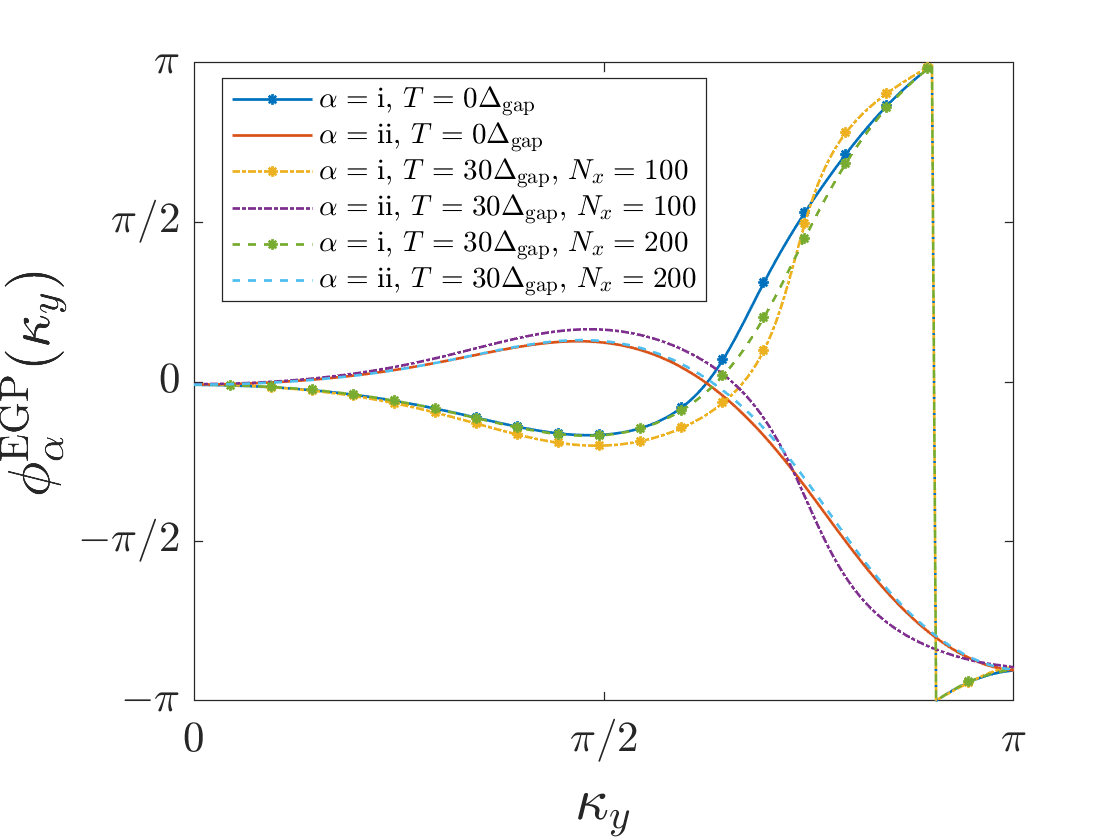}
\caption{TR EGP $\phi_\theta^\mathrm{EGP}(\kappa_y)=\phi_{\mathrm{i}}^\mathrm{EGP}(\kappa_y)-\phi_{\mathrm{ii}}^\mathrm{EGP}(\kappa_y)$ (top) and subband EGP $\phi_\alpha^\mathrm{EGP}(\kappa_y)$ for $\alpha = i,ii$ (bottom) of the Kane-Mele model. Parameter of the model are $\lambda_\textrm{SO}=0.06 t$, $\lambda_\textrm{R}=0.05 t$, and $\lambda_\textrm{v}=0.1 t$.}
\label{fig:TR_EGP}
\end{figure}

To illustrate our results we finally calculate the time-reversal EGP for a finite-temperature state of the KM model at half filling. In Fig. \ref{fig:TR_EGP} we have plotted the partial EGPs $\phi^\mathrm{EGP}_{i,ii}(\kappa_y)$ as well as their difference at a temperature much larger than the single-particle energy gap and compare them to the ground state values. One recognizes that the different EGPs approach the corresponding values at the ground state for increasing system size. Furthermore while there is no quantized winding of the partial EGPs, the time-reversal EGP has a quantized winding of $2\pi$ corresponding to the topologically non-trivial phase of the KM model.

\section{Summary and Conclusion}

In the present paper we have discussed the topological classification of Gaussian mixed states of TR symmetric band structures in $1+1$ and $2$ dimensions in terms of a generalized $\mathbb{Z}_2$ topological invariant. Gaussian mixed states are fully characterized by the single-particle correlation matrix, which defines a fictitious Hamiltonian. For the important class of thermal equilibrium states of non-interacting fermions, the latter is given by the system Hamiltonian itself multiplied by the inverse temperature. The generalized symmetries of this fictitious Hamiltonian under unitary and anti-unitary transformations provide a full topological classification according to the ten fundamental classes. For systems with broken TR symmetry the topological invariant, the Chern number, can be expressed in terms of an expectation value of a unitary operator. This formulation can straight-forwardly be extended to mixed states leading to the concept of the ensemble geometric phase \cite{Bardyn-PRX-2018,Linzner-PRB-2016,Wawer-PRB-2021}. For thermal states of TR symmetric Hamiltonians that can be smoothly deformed into a spin-conserving Hamiltonian without closing an energy gap, the difference of the Chern numbers of the two spin components is a suitable $\mathbb{Z}_2$ topological invariant which can be directly generalized to mixed states following the concept of \cite{Bardyn-PRX-2018}. We have shown that such a extension is also possible for general TR invariant band structures. To this end we generalized the formulation of the $\mathbb{Z}_2$ invariant in terms of the winding of the cTR polarization \cite{Grusdt-PRA-2014} to Gaussian mixed states. We showed that a similar mechanisms as discussed in \cite{Bardyn-PRX-2018} applies and leads to a reduction of the mixed-state $\mathbb{Z}_2$ index to the corresponding value in the ground state of the fictitious Hamiltonian. We illustrated our findings for thermal states of the Kane-Mele model. Our numerical simulations verified that the mixed-state topological index agrees with that in the ground state even for temperatures much above the single-particle energy gap.

\subsection*{acknowledgement}
Financial support from the DFG through SFB TR 185, project number 277625399  is gratefully acknowledged. We would like to thank Sebastian Diehl and Alexander Altland for stimulating discussions.


\end{document}